\documentclass{article}
\usepackage{spconf,amsmath,graphicx}
\usepackage{amsthm,amssymb}
\usepackage{mathrsfs}
\usepackage{algorithm}
\usepackage{algpseudocode} 
\usepackage{stfloats}
\usepackage{booktabs}
\usepackage{float}
\usepackage{amssymb}
\usepackage{multirow}
\usepackage{bbding}
\usepackage[table,xcdraw]{xcolor}
\usepackage[hyperfootnotes=true]{hyperref}

\title{M$^{3}$V: A multi-modal multi-view approach for Device-Directed Speech Detection}
\name{Anna Wang$^{\ast}$, Da Liu$^{\ast}$ \thanks{$^{\ast}$ Equal contribution}, Zhiyu Zhang, Shengqiang Liu, Jie Gao, Yali Li$^{\dagger}$ \thanks{$^{\dagger}$ Corresponding author: yali.li@nio.com} }
\address{NIO}
\begin{document}
\maketitle
\begin{abstract}
With the goal of more natural and human-like interaction with virtual voice assistants, recent research in the field has focused on full duplex interaction mode without relying on repeated wake-up words. 
This requires that in scenes with complex sound sources, the voice assistant must classify utterances as device-oriented or non-device-oriented. The dual-encoder structure, which is jointly modeled by text and speech, has become the paradigm of device-directed speech detection. 
However, in practice, these models often produce incorrect predictions for unaligned input pairs due to the unavoidable errors of automatic speech recognition (ASR). 
To address this challenge, we propose M$^{3}$V, a multi-modal multi-view approach for device-directed speech detection, which frames we frame the problem as a multi-view learning task that
introduces unimodal views and a text-audio alignment view in
the network besides the multi-modal. 
Experimental results show that M$^{3}$V significantly outperforms models trained using only single or multi-modality and surpasses human judgment performance on ASR error data for the first time.
\end{abstract}
\section{Introduction}
\label{sec:intro}

Recently, virtual assistants (VAs) are becoming more prominent in all aspects of our lives, the most representative of which include mobile phone intelligent assistant Siri, home smart speaker Alexa, and car intelligent voice assistant NOMI. 
Users typically trigger an interaction via a wake-up word, such as "Hi NOMI" or press a physical button directly on the smart device \cite{garg2022device}. 
However, such an interaction is unnatural from a human dialogue perspective because it does not conform to the habits of human voice interaction which does not rely on wake-up words or physical triggers all the time. 
In order to make the interaction more smooth and humanized, a lot of intelligent assistants have begun to explore a full duplex interaction mode without repeated wake-up words \cite{huang2019study}.

In these cases, an obvious challenge is that the scene sound source is complex. For example, in the scene of the in-vehicle intelligent assistant, the information received by the intelligent device may either come from the voice of the user to the device, the conversation information between the users, or the voice of the device itself. Therefore, it is necessary for the VAs to distinguish between device-directed and non-device-directed \cite{rudovic2022streaming}.

Many VAs address the above problems through multi-modal approaches, including voice and text.
Mallidi et al. \cite{mallidi2018device} use acoustics, ASR decoder, and ASR 1-best hypothesis to model a classifier to distinguish device-directed queries from background speech in the context of interactions with voice assistants.
Gillespie et al. \cite{gillespie2020improving} propose a system to combine acoustic features with word-level semantic lexical features to improve directedness classification.
Vilaysouk et al. \cite{vilaysouk2021improving} integrate ASR decoder-based features and word embeddings as additional inputs to the final classification stage of the model.
The dual-encoder structure, which is jointly modeled by text and speech, has become the paradigm of device-directed speech detection.

\begin{figure*}[htb]
\includegraphics[scale=0.35]{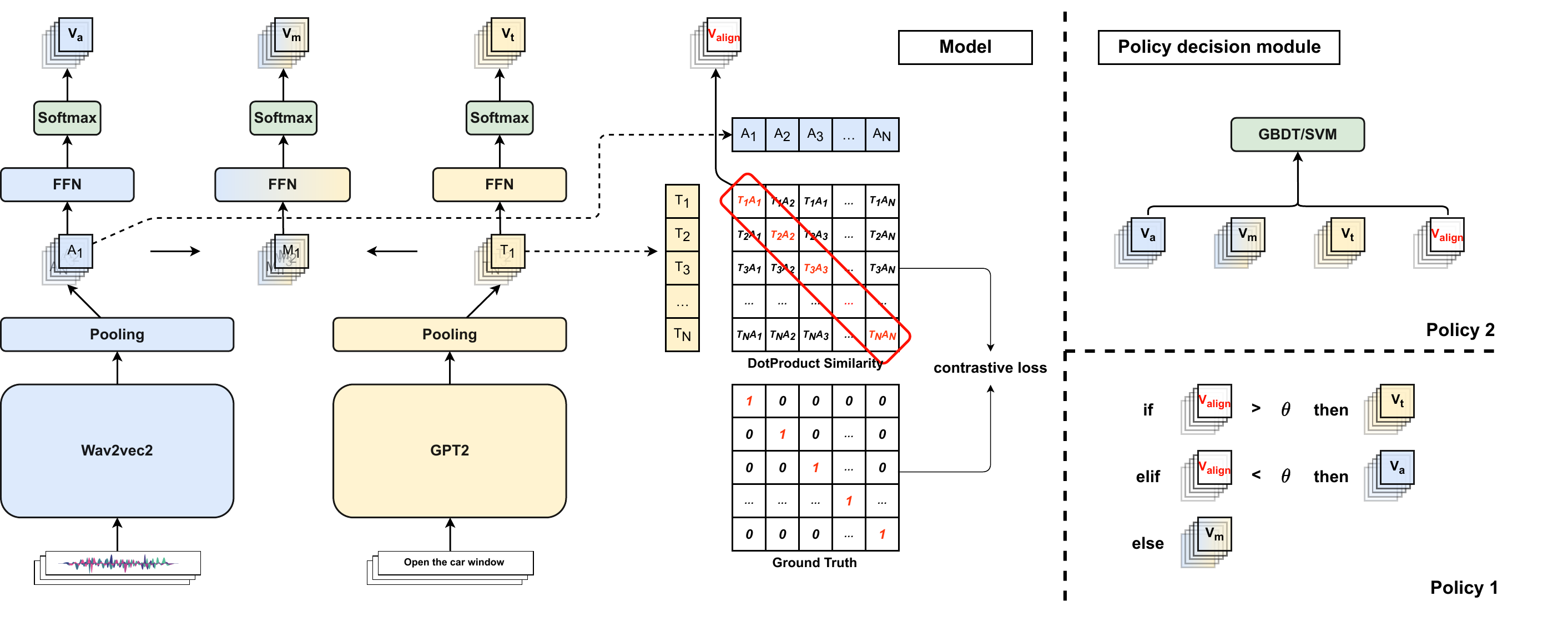}
\centering
\caption{Overview architecture of the M$^{3}$V. Given a text-audio pair as input, M$^{3}$V projects it to four views: unimodal-view ($V_{a}$, $V_{t}$), multi-modal-view($V_{m}$), aligned-view ($V_{algin}$). The predicted probabilities of these views are subject to arbitration by a policy decision module. }
\label{fig:M3V}
\end{figure*}

Despite the advances, a crucial limitation of the above models  is that they are mostly trained with supervised learning objectives only, where each text-audio pair is optimized with labeled ground truth and are thus never exposed to incorrect pairs during training, which hurts its generalization. This problem is even more severe in real-world VAs applications, as the input text is transformed by the ASR technique. 
Due to the error propagation of ASR, these models often produce incorrect predictions for unaligned input pairs. Additionally, the persistent modality gaps between heterogeneous modalities will further amplify the impact of ASR transcription errors, increasing the difficulty of analyzing multi-modal data.

Given the above problems, we propose a multi-modal multi-view approach (M$^{3}$V) for device-directed speech detection, which reduces the false dependence on ASR. 
Specifically, we use the pre-trained language model GPT2 \cite{lagler2013gpt2} to model text information and the pre-training model Wav2vec2 \cite{baevski2020wav2vec} to model audio information so as to make full use of the information of each mode.
Then, we frame the task as a multi-view learning problem to induce text-audio alignment information from a multi-modal model into our network using a contrastive loss function. 
In particular, we obtain multi-views, including unimodal information, multi-modal information, and alignment information.
These perspectives provide comprehensive and multi-faceted information for the task that can be combined for decision-making modules to eliminate the influence of ASR errors.  
Experimental results show that M$^{3}$V significantly outperforms models trained using only single or multi-modality and surpasses human judgment performance for the first time.

\section{Method}
\label{sec:format}
We propose a new framework that can judge whether it is device-oriented dialogue from multi-modal and multi-views.
The architecture of the model is shown in Figure 1. 
The model of M$^{3}$V is mainly divided into two stages, multi-modal learning based on device-oriented detection tasks and multi-view learning induces text-audio alignment information into the network using a contrastive loss function that obtains four views of information.
In addition, M$^{3}$V contains a policy decision module including two policies that can utilize four views of information for comprehensive evaluation of downstream tasks.
We describe each component in detail in this section.

\subsection{Multi-Modal Learning}

In multi-modal learning, we use text-audio pairs as the input of the model. Let the processed audio be $X_a$ and the text be represented by $X_t$. 
Given an audio-text pair of $N$ in a batch $\{X^{(i)}_a,X^{(i)}_t\}$ where $i\in\{1,2,...,N\}$, the problem is a binary classification problem to classify whether it’s device-directed or non-device-directed.

From the pairs, the audio and text are passed through an audio encoder and a text encoder, respectively. We choose a Wav2vec2 as the audio encoder which is proposed in \cite{baevski2020wav2vec} and a GPT2 \cite{lagler2013gpt2} as the text encoder which is a transformer-based language model.
Let ${f_{a}}$ represent the audio encoder and ${f_{t}}$ represent the text encoder. For an audio-text pair in a batch:

\begin{equation}\label{eq1}
A_{i} = \text{Pooling}({f_{a}}(\textbf{\footnotesize{X}}^{(i)}_{a}));T_{i} = \text{Pooling}({f_{t}}(\textbf{\footnotesize{X}}^{(i)}_{t})),~~i\in[1,N]
\end{equation}
\noindent
where $A_{i}\in\mathbb{R}^{d}$ 
is the audio representation of dimensionality ${d}$ 
and $T_{i}\in\mathbb{R}^{v}$ is the text representation of dimensionality ${t}$. 
A Pooling layer ($\mathit{e.g.,}$ mean-pooling) is applied to aggregate the frame-level features into an utterance-level representation.

By combining text representation and audio representation, we get multi-modal representation $M_{i}$.
\begin{equation}\label{eq2}
M_{i} = \text{Contact}({A_{i},T_{i}}),~~i\in[1,N]
\end{equation}

Then we input the text representation ${T_{i}}$, audio representation ${A_{i}}$, multi-modal representations ${M_{i}}$ into their respective feed forward network (FFN), and output multi-modal view result $V_{m}$ and unimodal view results $V_{a}$, ${V_{t}}$ after softmax:
\begin{equation}\label{eq3}
{V_{a}}= \text{FFN}_{a}({A_{i}});
{V_{t}}= \text{FFN}_{t}({T_{i}});
{V_{m}}= \text{FFN}_{m}({M_{i}})
\end{equation}

\subsection{Multi-view Learning}
In the previous section, we introduced the implementation of M$^{3}$V from a multi-modal view, but in practical applications, we found that it is limited.
Specifically, due to the unavoidable errors of ASR, these models often produce incorrect predictions for unaligned input pairs. 
In order to solve this problem, we frame the problem as a multi-view learning task that introduces unimodal views and a text-audio alignment view in the network besides the multi-modal view. 
As shown in Fig 1, the left of the model is a multi-modal learning network, which can produce multi-modal view and uni-modal view information. 
The right side is a contrastive model, which is capable of inferring whether text and audio are consistent by exploiting text and audio information that is only available during training.
For a contrastive model, we want the vectors of text-audio sources to be close to the uniform utterance. For this, we follow CLAP \cite{elizalde2022clap} and employ InfoNCE loss \cite{oord2018representation} to  measure dependencies between the audio and text modality. Consider the audio and text view of a min-batch as $A_{i}$  and $T{i}$, then the loss of contrastive model 
$\mathcal{L}_{c}$ can be defined as:

\begin{equation}\label{eq4}
\textbf{z}^{(i)}_{a} = \phi_{a}(A_{i}),  ~~\textbf{z}^{(i)}_{t} = \phi_{t}(T_{i}) \\
\end{equation}
\begin{equation}\label{eq5}
\mathcal{L}_{c}= -  \log  \frac{\mathrm{exp}(\mathrm{sim}(\textbf{z}^{(i)}_{a},\textbf{z}^{(i)}_{t})/\tau)}{\sum_{\textbf{z}^{(j)}_{t}\in\mathcal{B}}\mathrm{exp}(\mathrm{sim}(\textbf{z}^{(i)}_{a},\textbf{z}^{(j)}_{t})/\tau)} \\
\end{equation}

\noindent
where $\mathcal{B}=\{\textbf{z}^{(1)}_{\beta},\textbf{z}^{(2)}_{\beta},...,\textbf{z}^{(N)}_{\beta}\}$ ($\beta\in\{a,t\}$) is a set of hidden representations, which contains a positive sample $\textbf{z}^{(i)}_{\beta}$ and $\text{N-1}$ negative samples. $\phi_{a}$ is a learned projection function, $sim()$ is a similarity function ($\mathit{e.g.,}$ dot product), $\tau$ is a temperature parameter to scale the range of logits.
After the above learning, we have four views, which together affect the downstream task decision-making.
\subsection{Adaptive Learning}
We use two disjoint networks to share their learned information through the loss function \cite{lian2018speech}. 
The overall learning of the model is performed by minimizing:
\begin{equation}\label{eq6}
\mathcal{L} = \lambda \mathcal{L}_{a} + \gamma \mathcal{L}_{t} + \alpha \mathcal{L}_{m} +\beta \mathcal{L}_{c} 
\end{equation}
\noindent
where $\lambda$, $\gamma$, $\alpha$, $\beta$ are the interaction weights that determine the contribution of each regularization component to the overall loss $\mathcal{L}$. Each of these component losses using cross-entropy loss is responsible for achieving the desired subspace properties. Specially, we use the automatic weighted loss proposed by Lukas et al. \cite{liebel2018auxiliary}
\subsection{Policy Decision Module}
After the training of the above model, in the actual inference process, we obtain four probability scores, respectively from the four views.
To better adapt to specific downstream tasks, we propose three applied strategies to improve the overall accuracy of device-directed tasks. For final results, $True $ means device-directed and $False$ non-device-directed.

\noindent
\textbf{Policy 1:} In policy 1, we set five thresholds by evaluating the valid dataset, which acts on two policies respectively.
When the alignment score $S_{align}$ is greater than the high threshold of alignment $T_{align-high}$, the algorithm trusts the result of the text header.
When the alignment score $S_{align}$ is less than the low threshold of alignment $T_{align-low}$, the algorithm trusts the results of the audio header.
When the alignment score is between the high threshold $T_{align-high}$ and the low threshold $T_{align-low}$, the algorithm determines the final result according to the multi-modal fusion head.

\begin{algorithm}[th]
  \caption{The final decision of model and Policy 1.}
  \label{alg::conjugateGradient}
  \begin{algorithmic}[1]
    \Require
      The output of model $V_{align}$, $V_{audio}$, $V_{text}$, $V_{multi}$. And the threshold of above scores $T_{align-low}$, $T_{align-high}$, $T_{audio}$, $T_{text}$, $T_{multi}$
    \Ensure
      The label after Policy 1.
    \If {$V_{align} > T_{align-high}$}
        \State \Return {$V_{text} > T_{text}$};
    \ElsIf {$V_{align} < T_{align-low}$}
        \State \Return {$V_{audio} > T_{audio}$};
    \Else
        \State \Return {$V_{multi} > T_{multi}$};
    \EndIf
  \end{algorithmic}
\end{algorithm}
\begin{algorithm}[th]
  \caption{The final decision of model and Policy 2.}
  \label{alg::conjugateGradient}
  \begin{algorithmic}[2]
    \Require
      The output of model $V_{align} $,$V_{audio}$,$V_{text} $,$V_{multi}$ .And the threshold of fusion scores $T_{fusion}$
    \Ensure
      The label after Policy 2.
    \State $V_{fusion}$ = SVM($V_{align} $,$V_{audio}$,$V_{text} $,$V_{multi}$)
    \If {$V_{fusion} > T_{fusion}$}
        \State \Return  true;
    \Else
       \State \Return  false;
    \EndIf
  \end{algorithmic}
\end{algorithm}
\noindent
\textbf{Policy 2:} In policy 2, we will train a classifier, input the scores of four view results into the classifier, and determine whether it is a device-oriented dialogue by the prediction score by the classifier $V_{fusion}$ and the fusion threshold $T_{fusion}$. We tried several different machine learning classifiers, such as GDBT and SVM, and finally chose SVM with a better effect.
\section{Experiments}
\label{sec:experiments}
\subsection{Datasets}
We use real recordings of natural human interactions with in-vehicle virtual assistant NOMI for training and testing the models. 
The training dataset consists of 340 hours of audio data comprised of 500k utterances. 
The normal test data consists of 3.6 hours of audio data with 48k utterances. 
In addition, in order to prove that our model also has benefit compatibility in the case of weak ASR system capability, we selected a batch of ASR error data from real vehicle data, including 560 utterances with 55.60\% character error rate (CER). 
The model performance is evaluated in terms of equal error rate (EER) and accuracy.
\begin{table*}[htb]
\centering
\begin{tabular}{c|c|c|cccc|c|c}
\hline
\toprule
Modality & View & Model & Align ACC & Text ACC & Audio ACC & Merge ACC & EER &  $\bigtriangledown$ \\
\midrule Single & Single & GPT2 & —— & 91.41 & —— & —— & 10.61 & ——\\
\midrule Single & Single & Wav2vec2 & —— & —— & 92.41 & —— & 12.12 & ——\\
\hline
\midrule Multi & Single & \textit{Multi-modal} & —— & —— & —— & 95.27 & 6.59 & —— \\ 
\midrule Multi & Multi & M$^{3}$V Model& 85.48 & 91.43 & 95.62 & \textbf{96.27} & \textbf{4.94} &{\color[HTML]{009901} $\uparrow$ 1.65}\\
\bottomrule
\end{tabular}
\caption{Compare the performance of M$^{3}$V. The col with $\bigtriangledown$ means the improvements of our model compared to the Multi-modal in EER. A smaller ERR indicates better model performance.}
\label{multi-view}
\end{table*}
\subsection{Multi-Modal and Multi-View Experiments}
For comparison, we first train models with a single modality separately. We select the backbone on several encoders currently used in the text encoders including  GPT2 \cite{lagler2013gpt2}, BERT \cite{devlin2018bert}, Roberta \cite{liu2019roberta}, etc and audio encoders including Wav2vec2 \cite{baevski2020wav2vec}, Whisper \cite{radford2022robust}, SpeechTransformer \cite{dong2018speech}, etc. 
Finally, we choose the encoder with the highest accuracy in each modality of training data.
For audio modality, we use Wav2vec2 to model the sequence of speech frames. For text modality, we employ a GPT2 model which is based on the transformer. Then, we combine text and audio by training the \textit{Multi-modal} as the baseline of M$^{3}$V comparison.

As shown in Table 1, by combining speech and recognized text, the multi-modal approaches significantly boost unimodal approaches both text only and audio only, achieving 95.27\% accuracy and 6.59\% EER.
In comparison to multi-modal methods, the proposed approach outperforms the direct concatenation approaches, showing the advantage of learned alignment between speech and text. 

Our multi-view experiments use utterance-level representations to calculate the contrastive loss. Due to two disjoint networks sharing their learned information through the loss function, alignment information is learned in the multi-modal fusion head, resulting in a higher accuracy of 96.27\%. 
On this basis, the individual and aligned modal results are also output. Results from multiple perspectives will be used for downstream strategies.
\subsection{Policy decision experiments}
In the policy experiment evaluation module, we select 560 utterances with ASR errors to evaluate the generalization of the model in a challenging scenario. 
In order to quantify the improvement of the method, we first performed an accuracy evaluation of human performance. 
We selected five annotators to label these data and finally took the average as the accuracy of human performance.

As shown in Table 2,  the accuracy of manual annotation is 92.60\%. By combining speech and text, the multi-modal model is improved by 0.79\% compared to the manual annotation, and the accuracy of using M$^{3}$V model is improved by 2.40\%, reaching an accuracy of 95\% in the ASR error data. 
\begin{table}[htbp]
\centering
\begin{tabular}{c|c|c}
\hline
\toprule
Model & ACC & $\bigtriangledown$ \\
\midrule Human performance & 92.60 & ——\\ 
\midrule Multi-modal & 93.39 & {\color[HTML]{009901} $\uparrow$ 0.79}\\ 
\midrule M$^{3}$V Model  & 95.00 &  {\color[HTML]{009901} $\uparrow$ 2.40}\\
\midrule $+$ Policy \uppercase\expandafter{\romannumeral1} & \textbf{95.54} & {\color[HTML]{009901} $\uparrow$ 2.94} \\ 
\midrule $+$ Policy \uppercase\expandafter{\romannumeral2} & \textbf{95.71} & {\color[HTML]{009901} $\uparrow$ 3.11} \\ 
\bottomrule
\end{tabular}
\caption{Compare the performance of policies on the ASR error dataset.  The col with $\bigtriangledown$ means the model's improvements compared to manual annotation in accuracy.}
\label{policy results}
\end{table}

By feeding the multi-view results of the model into the policy decision module, the accuracy of the task is further significantly improved. Compared with manual annotation, the accuracy of Policy \uppercase\expandafter{\romannumeral1} is improved by 2.94\%, while Policy \uppercase\expandafter{\romannumeral2} is more significantly improved by 3.11\%, achieving the best effect on the ASR error dataset. The above experiments proved that the model plus strategy method could reduce the dependence on ASR error and improve model robustness. 
\section{CONCLUSION}
\label{conclusion}
In this paper, we propose M$^{3}$V, a multi-modal and multi-view approach for device-directed speech detection. 
Not only can the method learn the unimodal and multi-modal information after multi-modal fusion, but also it learns the alignment information between text-audio through disjoint networks. 
We show that M$^{3}$V outputs for downstream task decision-making and performs well across ASR error data, comparing with the unimodal and the direct concatenation multi-modal. 
This model achieves an accuracy of 96.41\% on the normal test set and 95.71\% on the ASR error test set. 
Especially, M$^{3}$V surpasses human judgment performance for the first time.
As a continuation of this work, we are considering incorporating more perspective information into the model, such as higher-order features in dialogue, to improve accuracy.

\vfill\pagebreak

\bibliographystyle{IEEEbib}
\bibliography{refs}

\begin{thebibliography}{10}

\bibitem{garg2022device}
Vineet Garg, Ognjen Rudovic, Pranay Dighe, Ahmed~H Abdelaziz, Erik Marchi,
  Saurabh Adya, Chandra Dhir, and Ahmed Tewfik,
\newblock ``Device-directed speech detection: Regularization via distillation
  for weakly-supervised models,''
\newblock {\em arXiv preprint arXiv:2203.15975}, 2022.

\bibitem{huang2019study}
Che-Wei Huang, Roland Maas, Sri~Harish Mallidi, and Bj{\"o}rn Hoffmeister,
\newblock ``A study for improving device-directed speech detection toward
  frictionless human-machine interaction.,''
\newblock in {\em INTERSPEECH}, 2019, pp. 3342--3346.

\bibitem{rudovic2022streaming}
Ognjen~Oggi Rudovic, Akanksha Bindal, Vineet Garg, Pramod Simha, Pranay Dighe,
  and Sachin Kajarekar,
\newblock ``Streaming on-device detection of device directed speech from voice
  and touch-based invocation,''
\newblock in {\em ICASSP 2022-2022 IEEE International Conference on Acoustics,
  Speech and Signal Processing (ICASSP)}. IEEE, 2022, pp. 491--495.

\bibitem{mallidi2018device}
Sri~Harish Mallidi, Roland Maas, Kyle Goehner, Ariya Rastrow, Spyros Matsoukas,
  and Bj{\"o}rn Hoffmeister,
\newblock ``Device-directed utterance detection,''
\newblock {\em Proc. Interspeech 2018}, pp. 1225--1228, 2018.

\bibitem{gillespie2020improving}
Kellen Gillespie, Ioannis~C Konstantakopoulos, Xingzhi Guo, Vishal~Thanvantri
  Vasudevan, and Abhinav Sethy,
\newblock ``Improving device directedness classification of utterances with
  semantic lexical features,''
\newblock in {\em ICASSP 2020-2020 IEEE International Conference on Acoustics,
  Speech and Signal Processing (ICASSP)}. IEEE, 2020, pp. 7859--7863.

\bibitem{vilaysouk2021improving}
Vilayphone Vilaysouk, Amr Nour-Eldin, and Dermot Connolly,
\newblock ``Improving identification of system-directed speech utterances by
  deep learning of asr-based word embeddings and confidence metrics,''
\newblock in {\em ICASSP 2021-2021 IEEE International Conference on Acoustics,
  Speech and Signal Processing (ICASSP)}. IEEE, 2021, pp. 6379--6382.

\bibitem{lagler2013gpt2}
Klemens Lagler, Michael Schindelegger, Johannes B{\"o}hm, Hana Kr{\'a}sn{\'a},
  and Tobias Nilsson,
\newblock ``Gpt2: Empirical slant delay model for radio space geodetic
  techniques,''
\newblock {\em Geophysical research letters}, vol. 40, no. 6, pp. 1069--1073,
  2013.

\bibitem{baevski2020wav2vec}
Alexei Baevski, Yuhao Zhou, Abdelrahman Mohamed, and Michael Auli,
\newblock ``wav2vec 2.0: A framework for self-supervised learning of speech
  representations,''
\newblock {\em Advances in Neural Information Processing Systems}, vol. 33, pp.
  12449--12460, 2020.

\bibitem{elizalde2022clap}
Benjamin Elizalde, Soham Deshmukh, Mahmoud~Al Ismail, and Huaming Wang,
\newblock ``Clap: Learning audio concepts from natural language supervision,''
\newblock {\em arXiv preprint arXiv:2206.04769}, 2022.

\bibitem{oord2018representation}
Aaron van~den Oord, Yazhe Li, and Oriol Vinyals,
\newblock ``Representation learning with contrastive predictive coding,''
\newblock {\em arXiv preprint arXiv:1807.03748}, 2018.

\bibitem{lian2018speech}
Zheng Lian, Ya~Li, Jianhua Tao, and Jian Huang,
\newblock ``Speech emotion recognition via contrastive loss under siamese
  networks,''
\newblock in {\em Proceedings of the Joint Workshop of the 4th Workshop on
  Affective Social Multimedia Computing and First Multi-Modal Affective
  Computing of Large-Scale Multimedia Data}, 2018, pp. 21--26.

\bibitem{liebel2018auxiliary}
Lukas Liebel and Marco K{\"o}rner,
\newblock ``Auxiliary tasks in multi-task learning,''
\newblock {\em arXiv preprint arXiv:1805.06334}, 2018.

\bibitem{devlin2018bert}
Jacob Devlin, Ming-Wei Chang, Kenton Lee, and Kristina Toutanova,
\newblock ``Bert: Pre-training of deep bidirectional transformers for language
  understanding,''
\newblock {\em arXiv preprint arXiv:1810.04805}, 2018.

\bibitem{liu2019roberta}
Yinhan Liu, Myle Ott, Naman Goyal, Jingfei Du, Mandar Joshi, Danqi Chen, Omer
  Levy, Mike Lewis, Luke Zettlemoyer, and Veselin Stoyanov,
\newblock ``Roberta: A robustly optimized bert pretraining approach,''
\newblock {\em arXiv preprint arXiv:1907.11692}, 2019.

\bibitem{radford2022robust}
Alec Radford, Jong~Wook Kim, Tao Xu, Greg Brockman, Christine McLeavey, and
  Ilya Sutskever,
\newblock ``Robust speech recognition via large-scale weak supervision,''
\newblock Tech. {R}ep., Technical report, OpenAI, 2022. URL https://cdn.
  openai. com/papers/whisper. pdf, 2022.

\bibitem{dong2018speech}
Linhao Dong, Shuang Xu, and Bo~Xu,
\newblock ``Speech-transformer: a no-recurrence sequence-to-sequence model for
  speech recognition,''
\newblock in {\em 2018 IEEE International Conference on Acoustics, Speech and
  Signal Processing (ICASSP)}. IEEE, 2018, pp. 5884--5888.

\end{thebibliography}

\end{document}